\documentclass[%
 reprint,
 amsmath,amssymb,
 aip,jcp,
 floatfix
]{revtex4-2}

\usepackage{graphicx}
\usepackage{dcolumn}
\usepackage{bm}
\usepackage{color}
\usepackage{tabularx}

\usepackage[english]{babel}
\usepackage{natbib}
\usepackage[mathlines]{lineno}

\begin{document}

%\preprint{APS/123-QED}

\title{A step in the direction of resolving the paradox of Perdew-Zunger self-interaction correction} 

\author{Rajendra R. Zope}
\email{rzope@utep.edu}
\author{Yoh Yamamoto}
\author{Carlos Diaz}
\author{Tunna Baruah}
\affiliation{Department of Physics, University of Texas at El Paso, El Paso, Texas 79968\\
}
\author{Juan E. Peralta}
\author{Koblar A. Jackson}
\affiliation{Physics Department and Science of Advanced Materials Program, Central Michigan University, Mt. Pleasant, Michigan 48859\\
}
\author{Biswajit Santra}
\author{John P. Perdew}
\affiliation{Physics Department, Temple University, Philadelphia, Pennsylvania 19122; Chemistry Department, Temple University, Philadelphia, Pennsylvania 19122\\
}

\date{\today}

\begin{abstract}
  Self-interaction (SI) error, which results when exchange-correlation contributions to the total energy are approximated, limits the reliability of many density functional approximations. The Perdew-Zunger SI correction (PZSIC), when applied in conjunction with the local spin density approximation (LSDA), improves the description of many properties, but overall, this improvement is limited. 
  Here we propose a modification to PZSIC that uses an iso-orbital indicator to identify  regions where local SI corrections should be applied. Using this local-scaling SIC (LSIC) approach with LSDA, we analyze predictions for a wide range of properties including, for atoms, total energies, ionization potentials, and electron affinities, and for molecules, atomization energies, dissociation energy curves, reaction energies, and reaction barrier heights. LSIC preserves the results of  PZSIC-LSDA for properties where it is successful and provides dramatic improvements for many  of the other properties studied. Atomization energies calculated using LSIC are better than those of the Perdew, Burke, and Ernzerhof (PBE) generalized gradient approximation (GGA) and close to those obtained with the  Strongly Constrained and Appropriately Normed (SCAN) meta-GGA.  
  LSIC also restores the  uniform gas limit for the exchange energy that is lost in PZSIC-LSDA. 
  Further performance improvements may be obtained by an appropriate combination or modification of the local scaling factor and the particular density functional approximation.

\end{abstract}

\maketitle

\section{\label{sec:introduction}Introduction}

 The Kohn-Sham (KS) formulation of density functional theory (DFT) has become the most popular approach for studying the electronic, structural and other properties of molecular and condensed systems \cite{PhysRev.140.A1133}. KS-DFT is a formally exact theory\cite{PhysRev.140.A1133,levy1979universal} to obtain the ground-state energy and electron density, but its practical realization requires an approximation to the exchange-correlation density functional. The enormous popularity of DFT is due to the combined appeal of sufficiently accurate density functional approximations (DFAs), favorable scaling with respect to the number of atoms, and numerically accurate and efficient implementations that have resulted in numerous easy-to-use codes. The local spin density approximation (LSDA)\cite{PhysRev.140.A1133,RevModPhys.61.689,jones2015density}, based on the uniform electron gas model, was an early and simple DFA. The success of LSDA in describing the electronic properties of solids made DFT popular in the physics community.  Careful analysis attributed this success to the spherical exchange-hole of LSDA being a good approximation to the spherical average of the exact exchange-hole and to the satisfaction of the exchange correlation hole sum rule \cite{PhysRevB.13.4274,1975SSCom.17.1425L,RevModPhys.61.689}.  Subsequent improvements beyond the LSDA were obtained \cite{becke1988density,PhysRevB.46.6671,PhysRevB.54.16533,PhysRevLett.77.3865,PKZB,perdew2001jacob,iikura2001long,PhysRevLett.91.146401,LEHTOLA20181,zhao2008density} by including information about the local electron density gradient in generalized gradient approximations (GGAs), and also the Laplacian and kinetic energy density, in meta-GGAs.  The non-empirical functionals among these are designed to satisfy various constraints and norms of the exact functional, 
 including the uniform electron gas limit \cite{doi:10.1063/1.1904565}. 
 
 Extensive work has shown that local and semi-local DFAs work well when the exact exchange-correlation hole density is localized around the electron, as is usually the case near equilibrium configurations in molecules and solids.  But these functionals can fail dramatically in stretched-bond situations such as in the transition states of chemical reactions and molecular dissociation \cite{doi:10.1063/1.1904565}, causing the underestimation of barrier heights in chemical reactions and the incorrect dissociation of radical and heteroatomic molecules. This failure can be traced to electron self-interaction errors (SIE) caused by the incomplete cancellation of the self-Coulomb energy with the approximate self-exchange-correlation energy for one electron densities.  This was  recognized long ago and  attempts to remove  SIE were 
 pursued \cite{lindgren1971statistical,perdew1979orbital, PhysRevA.15.2135,zunger1980self,gunnarsson1981self}.
 One widely-used approach to mitigating the effect of SIE, introduced by Becke, is 
 by combining Hartree-Fock exchange with semi-local functionals \cite{doi:10.1063/1.464304}. As the Hartree-Fock approximation is self-interaction free and introduces errors that are often of 
 opposite sign to those of semi-local functionals \cite{doi:10.1021/acs.jpclett.8b02417}, this approach can overcome a number of deficiencies of semi-local DFAs.
 The formal justification for such mixing can be obtained by an adiabatic connection \cite{PhysRevA.29.1648,PhysRevB.13.4274,1975SSCom.17.1425L}
 between the real interacting system and the non-interacting Kohn-Sham system. 
 Global hybrids \cite{doi:10.1063/1.464304}, local hybrids \cite{jaramillo2003local} and range-separated hybrids \cite{doi:10.1063/1.1383587},
 are all approximations that add Hartree-Fock exchange using various
 criteria. A majority of these functionals, however, still suffer from
 non-zero SIE. 
 
  A systematic procedure for eliminating one-electron self-interaction error 
 was given by Perdew and Zunger (PZ) in 1981 \cite{PhysRevB.23.5048}. In the PZ self-interaction correction (PZSIC) approach, 
 the SIE of a DFA is removed from the 
 total energy in  an orbital-by-orbital fashion by re-defining the energy as 
      
\begin{equation}\label{eq:DFASIC}
    E^{PZSIC-DFA} = E^{DFA}[\rho_{\uparrow},\rho_{\downarrow}]- \sum_i \left \{ U^{}[\rho_{i\sigma}]+E_{XC}^{DFA}[\rho_{i\sigma},0]\right\}.
\end{equation}
Here, $U[\rho_{i\sigma}]$ is the exact self-Coulomb energy and $E_{XC}^{DFA}[\rho_{i\sigma},0]$ is the approximate self-exchange and correlation energy.
PZSIC-DFA is exact for any one-electron density and gives no correction to the exact functional. 

One of the features of PZSIC is that $E^{PZSIC-DFA}$ is not invariant 
to the choice of orbitals used to represent the total electron density. Different orbitals that give the same total density yield different total energies so that finding the minimum energy formally requires searching over all sets of orbitals that span the correct density.
It can be shown that the variational minimum energy corresponds to $\rho_{i\sigma}=|\phi_{i\sigma}|^2$ for orbitals $\phi_{i\sigma}$ that satisfy the set of conditions known as the
Pederson or localization equations \cite{doi:10.1063/1.446959, doi:10.1063/1.448266},
\begin{equation}\label{eq:localization}
    \langle \phi_{i\sigma}|V_{i\sigma}^{SIC} - V_{j\sigma}^{SIC}| \phi_{j\sigma} \rangle = 0.
\end{equation}
In traditional PZSIC, a unitary transformation of the KS orbitals is performed to construct the local orbitals.
Optimizing the local orbitals to satisfy Eq.
(\ref{eq:localization}) requires tuning the O($N^2$) elements of the transformation matrix, which is computationally 
expensive. 

PZSIC provided a way to go beyond the LSDA,
but the computational difficulties mentioned above deterred practitioners from following this path\cite{Pederson_5scientific} 
and only a relatively few implementations of PZSIC
 have been reported \cite{ doi:10.1063/1.481421, doi:10.1063/1.1327269, doi:10.1063/1.1370527, harbola1996theoretical, doi:10.1063/1.1468640, doi:10.1021/jp014184v,PhysRevA.55.1765,doi:10.1080/00268970110111788, Polo2003, doi:10.1063/1.1630017, B311840A,doi:10.1063/1.1794633, doi:10.1063/1.1897378, doi:10.1063/1.2176608, zope1999atomic, doi:10.1063/1.2204599,doi:10.1002/jcc.10279,PhysRevA.45.101, PhysRevA.46.5453,lundin2001novel, PhysRevA.47.165,doi:10.1021/acs.jctc.6b00347,csonka1998inclusion,petit2014phase,kummel2008orbital,schmidt2014one,kao2017role,schwalbe2018fermi,jonsson2007accurate,rieger1995self,temmerman1999implementation,daene2009self,szotek1991self,messud2008time,messud2008improved,doi:10.1063/1.1926277,korzdorfer2008electrical,korzdorfer2008self,ciofini2005self}. 
 A review\cite{Pederson_5scientific} by Pederson and Perdew 
nicely summarizes this and related work. A handful of studies involved PZSIC combined with
semi-local approximations \cite{doi:10.1063/1.1794633,doi:10.1063/1.1897378,doi:10.1063/1.2176608, doi:10.1021/ct500637x, FLOSICSCANpaper, jonsson2007accurate}. 
These found that while PZSIC improves properties like the 
dissociation pathway of heteroatomic molecules,
it worsens the good description 
of semi-local functionals for near-equilibrium 
properties such as atomization energies, due to overcorrection \cite{JONSSON20151858, doi:10.1063/1.1897378}.  This has come to be known as the paradox of SIC \cite{PERDEW20151}.
A few approaches have been proposed to rectify this 
behavior based on scaling down the SIC contribution to the energy (second terms in the right hand side of Eq.~(\ref{eq:DFASIC})). Ref. [\citenum{PKZB}] proposed to use the ratio between the von Weizs\"{a}cker and the total kinetic energy densities to identify one- and two-electron regions for meta-GGAs, and Tsuneda \emph{et al.}\cite{doi:10.1002/jcc.10279} first proposed to use this ratio to identify one-electron regions where SIC is expected to be important. Ref. [\citenum{doi:10.1002/jcc.10279}] replaced the DFA energy density in these regions with an expression based on the exchange energy of hydrogenic orbitals.  Later, Vydrov \emph{et al.}\cite{doi:10.1063/1.2176608} used a selective orbital-by-orbital scaling down of the SIC contribution to the energy, and more recently,
J\'onsson \emph{et al.}\cite{doi:10.1063/1.4752229} proposed to globally reduce the SIC  energy by 50\%.  The J\'onsson group also pioneered\cite{PhysRevA.84.050501} the use of complex orbitals in PZSIC, which work well with PZSIC-PBE. 
The scaling approaches, which are discussed in more detail below, achieve success for selected properties, but, in general, they destroy the desirable $-1/r$ 
asymptotic form of the potential seen by an electron in a localized system such as a neutral atom in a PZSIC calculation \cite{doi:10.1063/1.2176608}. This unphysical behavior has important consequences for properties like charge transfer.

Considerable effort has been spent trying to understanding the origin of the PZSIC paradox.  A recent study found that PZSIC
raises the total energy as the nodality of the valence local orbitals increases from atoms to molecules to transition states \cite{doi:10.1063/1.5087065}. More recently, it was shown that, 
unlike the non-empirical semi-local functionals, PZSIC violates the uniform electron gas norm for the exchange and correlation energies \cite{doi:10.1063/1.5090534}.  The implication of this is that adding PZSIC breaks the correct behavior of these functionals for slowly-varying densities.

In this work, we propose an approach that adjusts the PZSIC correction
locally, that is, at each point in space, by adjusting the magnitude of the correction using an iso-orbital indicator. We call this approach local-scaling SIC (LSIC). It is implemented 
in the FLOSIC code \cite{FLOSICcode,FLOSICcodep} and applied perturbatively to self-consistent PZSIC solutions obtained using the Fermi-L\"owdin orbital SIC (FLO-SIC) method \cite{doi:10.1063/1.4869581, PEDERSON2015153}.  As discussed further below, the method applies SIC at full strength for a density with a single-orbital character and turns it off for a uniform density.
We assessed the predictions of this approach for a number of properties including, for atoms:  total energies, ionization potentials, and electron affinities, and for molecules: atomization energies,  reaction energies and dissociation energy curves, and reaction barrier heights.  We find significant improvement for properties that PZSIC typically worsens, while retaining the successful predictions of PZSIC in situations where removing SIE is critical. The proposed LSIC method, unlike semi-local functionals and most earlier PZSIC implementations,  provides a good  description of both near-equilibrium properties and properties associated with stretched-bond situations.  
LSIC thus appears to resolve the paradox of PZSIC and opens the door to designing universally accurate DFAs.

\section{\label{sec:theory}Theory and Computational Details}

   The application of PZSIC worsens the quality of equilibrium properties when used
   with semilocal functionals \cite{doi:10.1063/1.1794633,doi:10.1063/1.1897378,doi:10.1063/1.2176608,FLOSICSCANpaper,PhysRevA.84.050501,PhysRevA.100.012505,dipolepaper}. 
    Attempts have been made to restore the accuracy 
of semi-local functionals used in combination with  PZSIC by reducing the size of the corrections.
For example, J\'onsson and coworkers used a scaled-down version of PZSIC in which the SIC correction 
is reduced by 50\% \cite{doi:10.1063/1.4752229}.  Such a diminished correction, when applied with the PBE functional, resulted in overall
improvement of atomization energies but significant absolute errors still remained. 
Instead of using a fixed constant scaling factor, Vydrov and coworkers \cite{doi:10.1063/1.2176608}
had earlier proposed setting a scaling factor for each local orbital $i$ in the following way:
\begin{equation}\label{eq:factorscuseria}
    X_{i\sigma}^k = \int \left(\frac{\tau_\sigma^W}{\tau_\sigma}\right)^k \rho_{i\sigma}(\vec{r}) d\vec{r}.
\end{equation}
Here, $\tau_\sigma^W(\vec{r})$  is the von Weisz\"acker kinetic energy density
and $\tau(\vec{r})$ is the Kohn-Sham kinetic energy density.
This scaling factor is subsequently used to attenuate the Coulomb and exchange-correlation parts of the SIC as follows:
\begin{equation}\label{eq:scalingscuseria}
    E^{scaled-SIC} = - \sum_{i\sigma}^{occ} X_{i\sigma}^k (U[\rho_{i\sigma}]+E_{XC}^{DFA}[\rho_{i\sigma},0]).
\end{equation}
We shall refer to this approach as exterior orbital scaling.  Like PZSIC, and unlike the 50\% global scaling, this approach is exact for all one-electron densities and, with $k\ge 1$, for all uniform densities.  

Vydrov \emph{et al.} noted that while increasing $k$ above zero satisfies some additional exact constraints, the correct
$-1/r$ asymptotic behavior of the one electron potential is lost if $k>0$.
Vydrov and Scuseria also proposed a simpler method \cite{doi:10.1063/1.2204599} of 
moderating SIC with a scaling factor given as
\begin{equation}
    W_{i\sigma}^m = \int \left(\frac {\rho_{i\sigma}}{\rho_{\sigma}}\right)^m \rho_{i\sigma} dr = \int \frac{\rho_{i\sigma}^{m+1}}{\rho_\sigma^m} d\vec{r}.
\end{equation}
This factor depends on the ratio of overlaps of orbital density $\rho_{i\sigma}$ and 
the total density $\rho_{\sigma}$ for a given spin $\sigma$. 
The authors noted that the
SIC-PBE with $m=1$ performs consistently well for the benchmark tests, but a larger value 
of $m$, such as $m=4$, is needed for SIC-LSDA.

PZSIC improves results where semi-local functionals
fail drastically on account of SIE \cite{doi:10.1021/acs.jpca.8b09940,doi:10.1063/1.2176608, doi:10.1063/1.5087065}, 
but it overcorrects and
worsens the description of near-equilibrium properties such as molecular atomization energies.
Based on this observation, we propose a modification of PZSIC in such a way that
the  self-interaction correction is enforced only where it is necessary. 
This can be done {\it locally}, or point-wise in space, that is, it can 
be applied only
in the regions where self-interaction is expected to be strong. 
Tsuneda and coworkers\cite{doi:10.1002/jcc.10279} defined these to be regions where the density has one-electron character and they used the ratio 
 $z_\sigma(\vec{r})=\tau_\sigma^W(\vec{r})/\tau_\sigma(\vec{r})$  to identify these regions. 
 Here the non-interacting (Kohn-Sham) kinetic energy density  $\tau_\sigma$ for a 
 spin $\sigma$ is given as,
\begin{equation}
    \tau_\sigma(\vec{r}) = \frac 1 2 \sum_i |\vec{\nabla} \psi_{i\sigma}(\vec{r})|^2,
\end{equation}
and $\tau_\sigma^W$ is given as
\begin{equation}
    \tau_\sigma^W(\vec{r}) = \frac{|\vec{\nabla} \rho_\sigma(\vec{r})|^2}{8 \rho_\sigma(\vec{r})}.
\end{equation}
Since $\tau_\sigma^W$ is the single orbital limit of $\tau_\sigma$ and  vanishes for a uniform density,  $z_\sigma(\vec{r})$
varies between zero and one, with zero corresponding to uniform densities and one to one-electron densities. In their regional SIC scheme, Tsuneda and
coworkers\cite{doi:10.1002/jcc.10279} used this ratio  to replace the conventional DFT expression for the exchange and correlation potential in regions where $z$ is close to one by a simple model expression intended to mimic the exchange potenial of a single hydrogenic orbital.   They used their scheme to study reaction barriers, where they
found significant improvement over conventional DFT calculations. Following Tsuneda \emph{et al.}, we propose the following modification
to the PZSIC energy expression:
\begin{equation}\label{eq:LSIC-DFA}
    E_{XC}^{LSIC-DFA} = E_{XC}^{DFA}[\rho_{\uparrow},\rho_{\downarrow}]
                       - \sum_{i,\sigma}^{occ}
     \left \{ U^{LSIC}[\rho_{i,\sigma}]
     + E_{XC}^{LSIC}[\rho_{i,\sigma},0] \right \}
\end{equation}
where 
\begin{equation}\label{eq:lsiccoul}
     U^{LSIC}[\rho_{i,\sigma}] = 
       \frac{1}{2} \int d\vec{r}\,
       \{z_{\sigma}(\vec{r})\}^k \,
       \rho_{i,\sigma}(\vec{r}) 
       \int d\vec{r'}\, \frac{\rho_{i,\sigma}(\vec{r'})}{\vert \vec{r}-\vec{r'}\vert}
\end{equation}
and
\begin{equation}\label{eq:lsicxc} 
     E_{XC}^{LSIC}[\rho_{i,\sigma},0]
     =  \int d\vec{r}\,  \{z_{\sigma}(\vec{r})\}^k
     \rho_{i,\sigma} (\vec{r}) \epsilon_{XC}^{DFA}([\rho_{i,\sigma},0],\vec{r}). 
\end{equation}

The LSIC-DFA of Eq. [\ref{eq:LSIC-DFA}] recovers the PZSIC (Eq. [\ref{eq:DFASIC}]) for $k=0$. The $k\rightarrow\infty$ limit, on the other hand, zeroes out the SIC and reduces thereby to a standard DFA, except in fully one-electron regions.
In the present work we use $k=1$. This is the simplest choice of scaling factor based on $z_{\sigma}$.  It smoothly interpolates between uniform density regions and one-electron regions.
In the rest of this section we provide the computational details.

We implemented the LSIC approach in the FLOSIC code which is based on 
the UTEP-NRLMOL \cite{FLOSICcode,FLOSICcodep} software package.  UTEP-NRLMOL is a modern version
of the Gaussian-orbital-based NRLMOL code \cite{doi:10.1002/(SICI)1521-3951(200001)217:1<197::AID-PSSB197>3.0.CO;2-B, PhysRevB.41.7453, PhysRevB.42.3276}. 
We use the NRLMOL default basis
sets \cite{PhysRevA.60.2840} that are of approximately quadruple zeta quality \cite{doi:10.1002/jcc.25586}.  For a better description of atomic anions, we added long range s, p, and d single Gaussian orbitals to the default NRLMOL basis set.  The exponents for the additional functions were obtained from the relation $\beta(N+1) =  \frac{\beta(N)}{\beta(N-1)} \beta(N)$ where $\beta(N)$ is the exponent of $N$-th Gaussian in the basis for a given atom.
NRLMOL's variational integration mesh \cite{PhysRevB.41.7453} adapts to the range of 
basis functions so that integrals are computed to a specified accuracy. 

  We use the Fermi-L\"owdin Orbital Self-Interaction Correction (FLO-SIC) approach proposed by Pederson \textit{et al.} \cite{doi:10.1063/1.4869581, PEDERSON2015153} to implement the PZSIC and LSIC methods.
In FLO-SIC, the local orbitals used to evaluate the PZSIC total energy are based on Fermi orbitals constructed from the density with parameters known as
Fermi orbital descriptors (FODs), $M$ positions in 3-dimensional space for $M$ occupied orbitals. Using these FODs, one can write the Fermi orbitals as
\begin{equation}\label{eq:fods}
    F_{i \sigma}(\vec{r}) = \sum_j^M \frac {\psi_{j \sigma}^*(\vec{a}_{i \sigma}) \psi_{j \sigma}(\vec{r}) } {\sqrt{\rho_{\sigma}(\vec{a}_{i \sigma})}} \,,
\end{equation}
where $\vec{a}_{i \sigma}$ is the FOD position for orbital $i$ of spin $\sigma$, $\rho_{\sigma}$ is the electron spin density, and $\psi_{j \sigma}$ is one of the $M$ occupied orbitals.  Since Fermi orbitals are generally not orthogonal,  L\"owdin orthogonalization is performed to transform the {$F_i$} into the orthonormal local orbital {$\phi_{i,\sigma}$}.  
The FLO-SIC approach is unitarily invariant because any set of orbitals spanning the occupied orbital space can be used in Eq. [\ref{eq:fods}]  
to generate the Fermi orbitals.  To minimize the PZSIC energy, the $3N$ FOD positions must be optimized.  This is done using gradients of the energy with respect to the FOD positions \cite{doi:10.1063/1.4907592} in a manner analogous to a molecular geometry optimization. This is a computationally simpler process than determining the O($N^2$) parameters required to define the local orbital transformation in
traditional PZSIC. 
We follow the self-consistency procedure of Ref. [\citenum{PhysRevA.95.052505}].  Iteration averaging was performed for the DFA potentials, using either Broyden mixing or simple mixing scheme to accelerate convergence. A self-consistency convergence tolerance of $10^{-6}$ Ha in the total energy was used for all calculations. For PZSIC calculations using FLOSIC, an FOD force tolerance of $10^{-3}$ Ha/Bohr was used to ensure optimized FOD positions.  LSIC-LSDA total energies (Eq. (\ref{eq:LSIC-DFA})) were computed using the corresponding self-consistent PZSIC-LSDA density and optimized local orbitals. 
Both LSIC and FLO-SIC calculations have similar computational costs. The only additional quantity needed for LSIC is the scaling factor, which requires the evaluation of the kinetic energy densities whose computational cost is negligible.
The FLO-SIC methodology has been recently employed to study atomic energies \cite{doi:10.1063/1.4996498}, atomic polarizabilities \cite{PhysRevA.100.012505}, and magnetic exchange couplings \cite{doi:10.1063/1.5050809}.

\section{Results}\label{sec:results}
%%%%%%%%%%%%%%%%%%%%%%%%%%%%%%%%%%%%%%%%%%%%%%%%%%%%%%%%%%%%%%%%%%
\subsection{Atoms}
\subsubsection{Total Energy}

\begin{figure}
\includegraphics[width=\columnwidth]{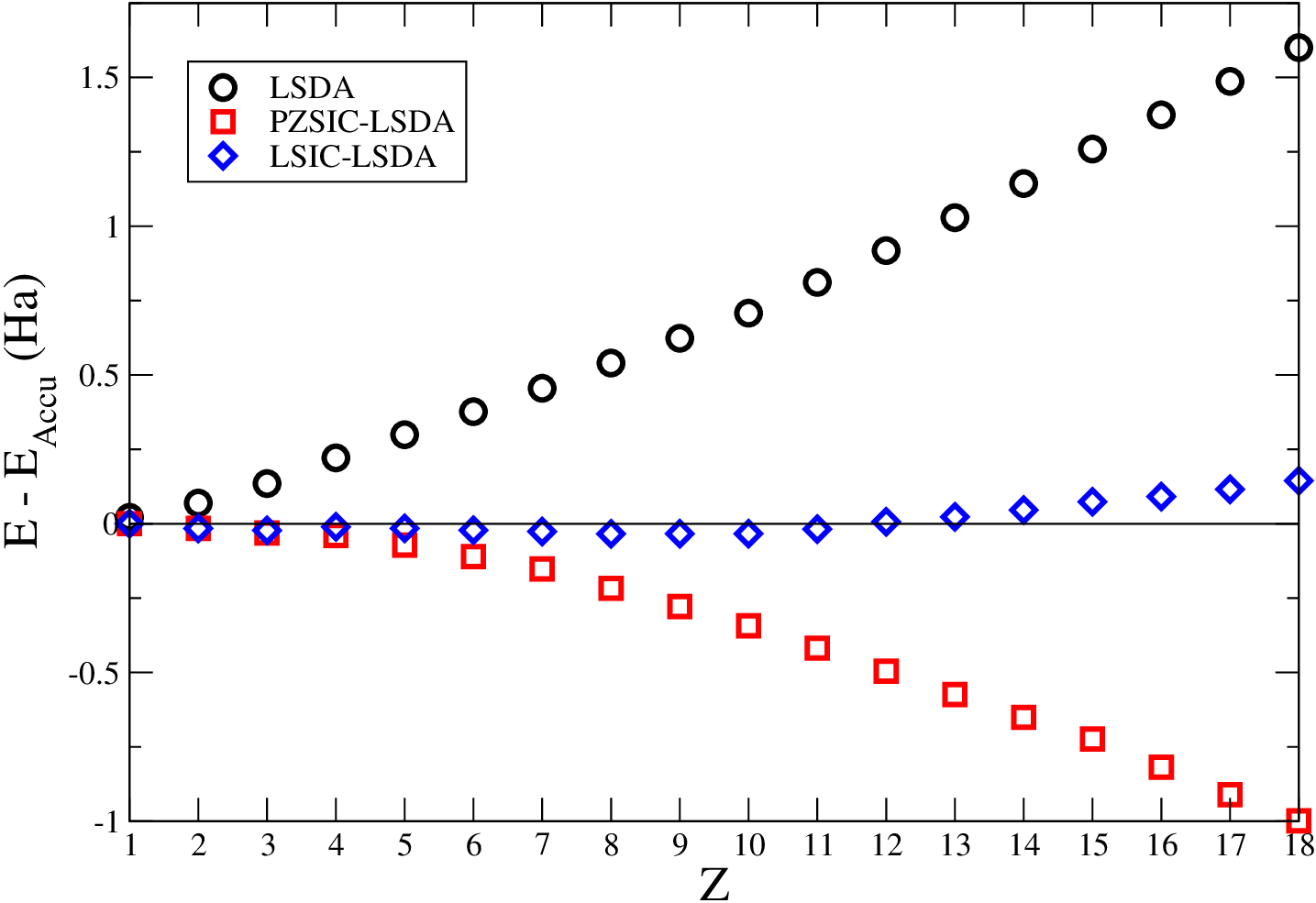}
\caption{\label{figatoms} LSDA, PZSIC-LSDA, and the LSIC-LSDA total energies of atoms Z $=1-18$, relative to reference energies ($E_{accu}$) from Ref. [\citenum{PhysRevA.47.3649}].}
\end{figure}

We compared the total energy $E$ of atoms for atomic numbers Z$=1-18$ computed using different 
methods against the accurate non-relativistic energies ($E_{accu}$) reported
by Chakravorty \textit{et al.} \cite{PhysRevA.47.3649}.
The total energy differences $E - E_{accu}$ are shown in Fig. \ref{figatoms} for LSDA, PZSIC-LSDA, and LSIC-LSDA. In 
general, LSDA overestimates the total energies and applying SIC 
shifts the energies in the proper direction.  But the corrections are too large and PZSIC-LSDA predicts atomic energies that are too low.  The LSIC-LSDA 
total energies, by contrast, are very close to the reference energies. LSIC reduces the mean absolute errors (MAE) in the energies by an order of magnitude compared to PZSIC-LSDA.
The MAE are 0.726, 0.381, and 0.041 Ha for LSDA, PZSIC-LSDA, and LSIC-LSDA, respectively.
Results for atomic total energies for a variety of methods are summarized in Table \ref{tabatoms}. The LSIC-LSDA results are better than PBE, but slightly worse than SCAN.

\begin{table}
\caption{\label{tabatoms}%
MAE of atom total energies Z=1--18 (Ha)
}
\footnotetext{Reference \cite{FLOSICSCANpaper}.}
\begin{ruledtabular}
\begin{tabular}{lc}
\textrm{Method}&
\textrm{MAE}\\
\colrule
LSDA$^\text{a}$ &  0.726  \\
PBE$^\text{a}$ & 0.083 \\ 
SCAN$^\text{a}$ & 0.019 \\ 
PZSIC-LSDA$^\text{a}$ & 0.381  \\
PZSIC-PBE$^\text{a}$ & 0.159  \\
PZSIC-SCAN$^\text{a}$ & 0.147 \\
LSIC-LSDA & 0.041  \\ 
\end{tabular}
\end{ruledtabular}
\end{table}

%%%%%%%%%%%%%%%%%%%%%%%%%%%%%%%%%%%%%%%%%%%%%%%%%%%%%%%%%%%%%%%%%%%
\subsubsection{Ionization potential}
\begin{figure}
\includegraphics[width=\columnwidth]{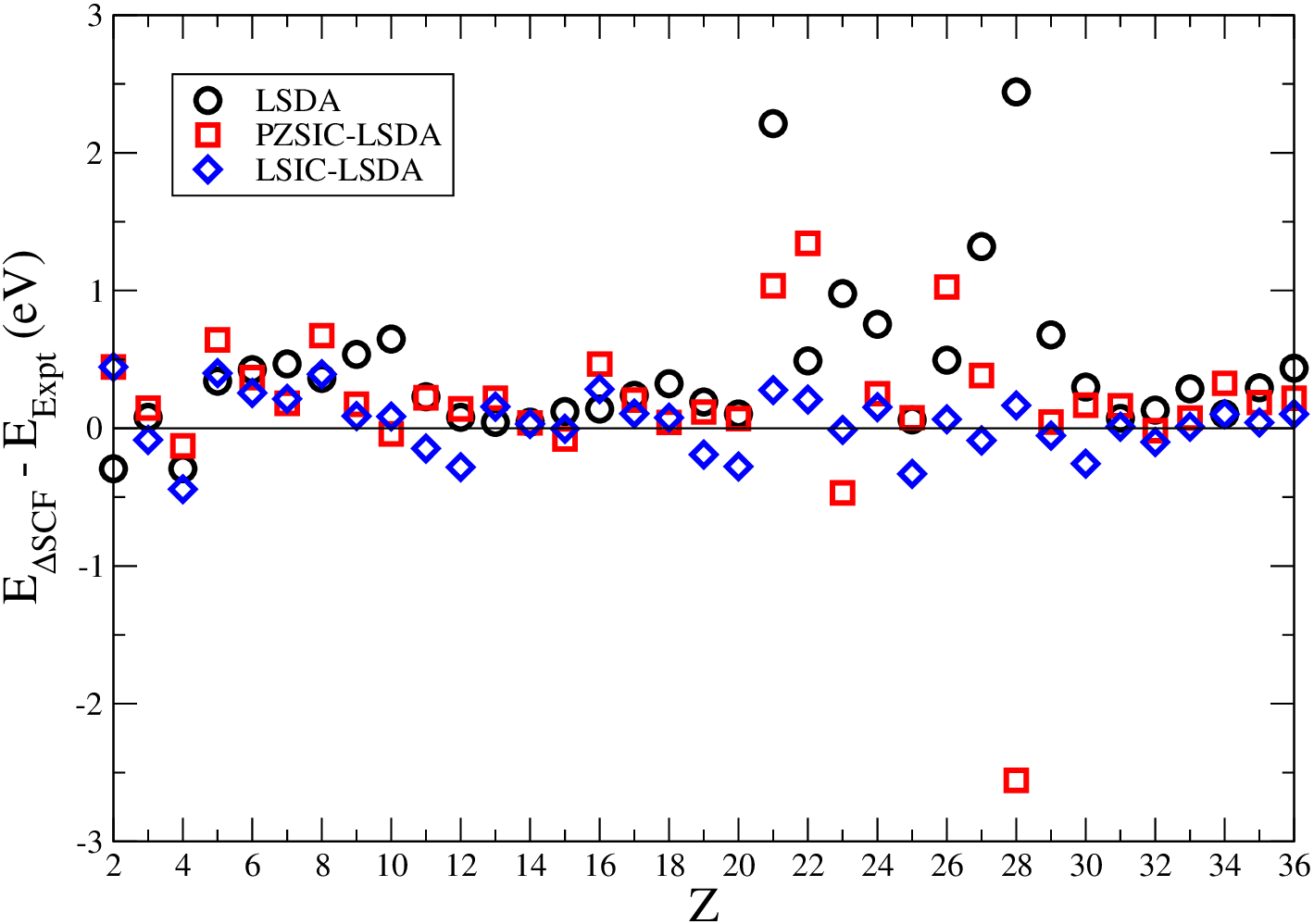}
\caption{\label{figip} Ionization potential of atoms Z $=2-36$. LSDA, PZSIC-LSDA, and the LSIC-LSDA are shown.}
\end{figure}

Since the ionization potential is the energy required to remove an electron from 
the outermost orbital, this quantity is  sensitive to 
the asymptotic behavior of the potential and can be expected to be affected by SIC, especially for large systems.  
We calculated the ionization potential (IP) for the atoms He--Kr using the $\Delta$SCF approach 
\begin{equation}
    E_{IP} = E_{cation} - E_{neutral}.
\end{equation}
Fig. \ref{figip} shows a comparison of calculated IPs against experimental values for LSDA, PZSIC-LSDA, and LSIC-LSDA.  The MAEs are presented in Table \ref{tabip}, along with results for other methods. From LSDA to PZSIC, the IPs improve noticeably, with the MAE dropping from 0.458 to 0.364 eV.   
LSIC further improves the IPs, reducing the MAE down to 0.170 eV.  Because the LSIC-LSDA total energies for the neutral atoms are very close to the reference energies, the accurate IP values imply that the LSIC-LSDA cation energies are also quite accurate.
In Table \ref{tabip} we show the results for the atoms from He--Ar and for all atoms in separate columns, to distinguish the performance for light versus heavy atoms. PBE and SCAN perform well for the lighter atoms, but less so for the heavier ones. LSIC-LSDA, on the other hand, performs equally well for all atoms.  LSIC-LSDA performs better than both PBE and SCAN (MAE 0.253 and 0.273 eV, respectively) for the 35 IPs.

\begin{table}
\caption{\label{tabip}
MAE of $\Delta$SCF ionization potentials (eV)}
\begin{ruledtabular}
\begin{tabular}{lcc}
\textrm{Method}&
\textrm{Z=2--18 (17 IPs) MAE} &
\textrm{Z=2--36 (35 IPs) MAE}\\
\colrule
LSDA  &0.275& 0.458 \\ 
PBE   &0.159& 0.253 \\ 
SCAN  &0.175& 0.273 \\ 
PZSIC-LSDA  &0.248& 0.364 \\ 
PZSIC-PBE   &0.405& 0.464 \\ 
PZSIC-SCAN  &0.274& 0.259 \\ 
LSIC-LSDA       &0.206& 0.170 \\ 
\end{tabular}
\end{ruledtabular}
\end{table}

%%%%%%%%%%%%%%%%%%%%%%%%%%%%%%%%%%%%%%%%%%%%%%%%%%%%%%%%%%%%%%%%%%%%%%%%%%%%%%%%%%
\subsubsection{Electron affinity}
\begin{figure}
\includegraphics[width=\columnwidth]{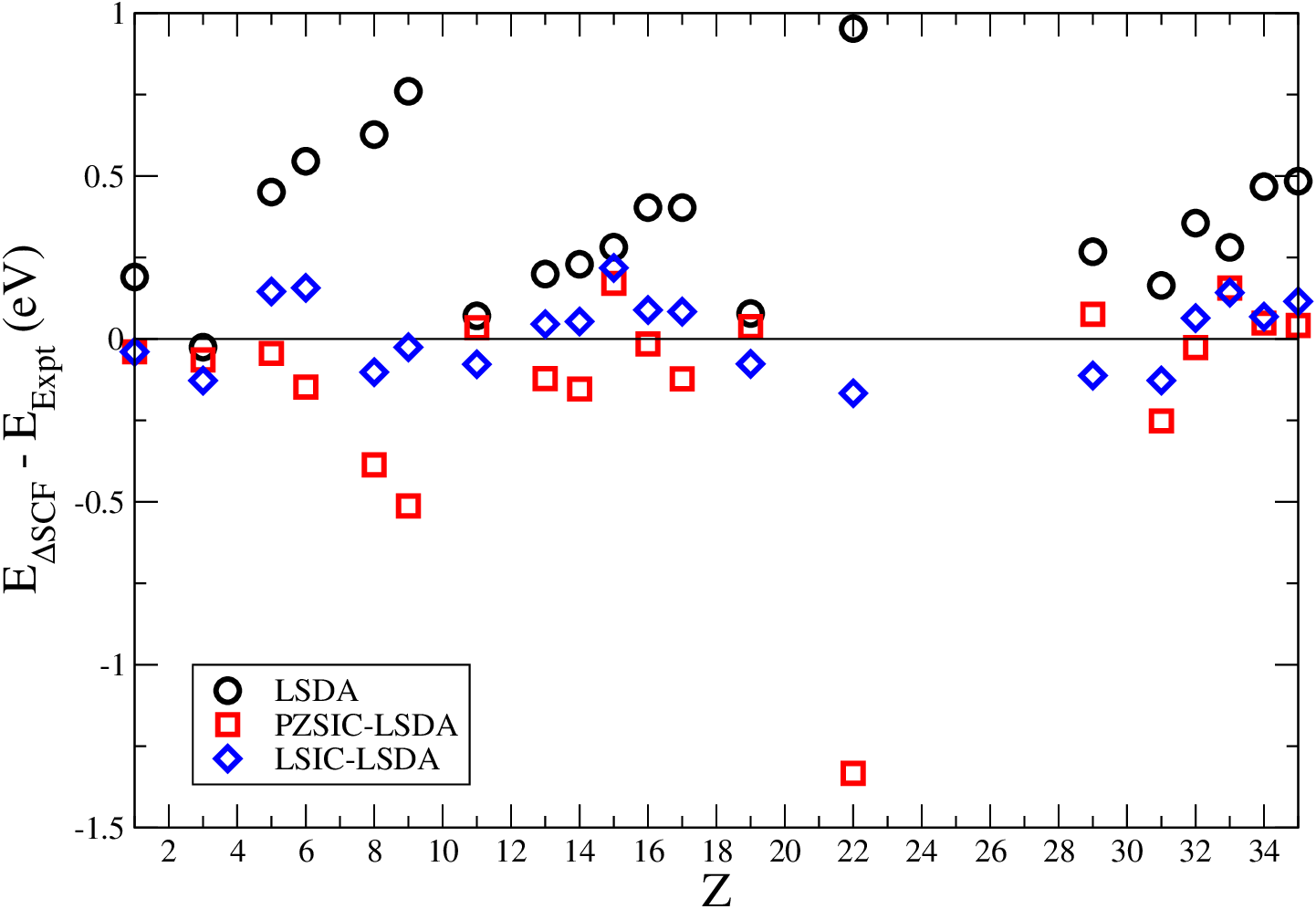}
\caption{\label{figea35} Electron affinity of atoms Z $=1-35$. LSDA, PZSIC-LSDA, and the LSIC-LSDA are shown where experimental values are reported.}
\end{figure}

The electron affinities (EA) of atoms from  H to Br were also investigated. 
As in the case of the IPs, the EAs were calculated using the $\Delta$SCF method, taking the total energy differences of the neutral atoms and their anions. 
We considered the twenty atoms in the first three rows (H, Li, B, C, O, F, Na, Al, Si, P, S, Cl, K, Ti, Cu, Ga, Ge, As, Se, and Br) with stable anions and for which experimental EAs are available \cite{NIST_CCCBD}. Fig. \ref{figea35} shows the results for each atom for LSDA, PZSIC-LSDA, and LSIC-LSDA.  In Table \ref{tabea} we again analyze the performance of various methods, dividing the results into two groups, the first containing the 12 EAs corresponding to first and second row atoms and the second containing all 20 EAs.  The MAE relative to experiment are shown in the table. 
We note that although the $\Delta$SCF approach yields positive EAs for the DFAs, the eigenvalue corresponding to the added electron becomes positive in all DFA anion calculations, indicating that the extra electron is not actually bound in the complete basis set limit.  This problem is due to the incorrect asymptotic form of the potential in the DFA calculations.  SIC rectifies this \cite{PhysRevB.23.5048}, leading to bound states for the HOMO in the anions. Nevertheless, we include the EAs of DFA calculations based on the $\Delta$SCF approach in Table \ref{tabea} for  comparison purposes. We note that these results compare well with PZSIC results of Vydrov and Scuseria \cite{doi:10.1063/1.2176608}. 

Overall, LSDA overestimates the values of the EAs, whereas PZSIC-LSDA tends to underestimate them, particularly for O, F, and Ti. The LSIC-LSDA method improves the EA values so that they consistently fall within $\pm$0.2 eV of the experimental values. The MAE of 20 EAs is reduced from 0.362 eV for LSDA, to 0.189 eV for PZSIC-LSDA, to 0.102 eV for LSIC-LSDA.  

\begin{table}
\caption{\label{tabea}%
MAE of $\Delta$SCF electron affinities (eV)}
\footnotetext{Reference \cite{FLOSICSCANpaper}.}
\footnotetext{DFA results are based on total energies. The eigenvalue of the extra electron becomes positive.}
\begin{ruledtabular}
\begin{tabular}{lccc}
\textrm{Method}&
\textrm{12 EAs MAE }&
\textrm{20 EAs MAE }\\
\colrule
LSDA$^\text{a,b}$ & 	   0.349   &   0.362	\\
PBE$^\text{a,b}$  &   0.167    &   0.172 \\
SCAN$^\text{a,b}$ &   0.115   &   0.148\\
PZSIC-LSDA$^\text{a}$ &    0.151   &  0.189 \\
PZSIC-PBE$^\text{a}$  &    0.534   &  0.531 \\
PZSIC-SCAN$^\text{a}$ &   0.364   &  0.341\\
LSIC-LSDA &  0.097   & 0.102 \\ 	
\end{tabular}
\end{ruledtabular}
\end{table}

%%%%%%%%%%%%%%%%%%%%%%%%%%%%%%%%%%%%%%%%%%%%%%%%%%%%%%%%%%%%%%%%%%%%%%%%%%%%%%%%%%%
\subsection{Atomization energy}
The atomization energy ($AE$) of a molecule is defined as
\begin{equation}
    AE =\sum_i^{N_{atom}} E_i - E_{mol} > 0,
\end{equation}
where $E_i$ is the energy of an individual atom and $E_{mol}$ is the energy of the molecule.
We computed $AE$s for a diverse set consisting of 37 molecules. The majority of the molecules were taken from the G2/97 test set \cite{doi:10.1063/1.460205}. We also included the six systems from the AE6 test set \cite{doi:10.1021/jp035287b}.
All molecular geometries were taken from Ref.  [\citenum{NIST_CCCBD}] (B3LYP and the 6-31G(2df,p) basis) except O$_2$, CO, CO$_2$, C$_2$H$_2$, Li$_2$, CH$_4$, NH$_3$, and H$_2$O, which were obtained using PBE and the default NRLMOL  basis set.

\begin{figure}
\includegraphics[width=\columnwidth]{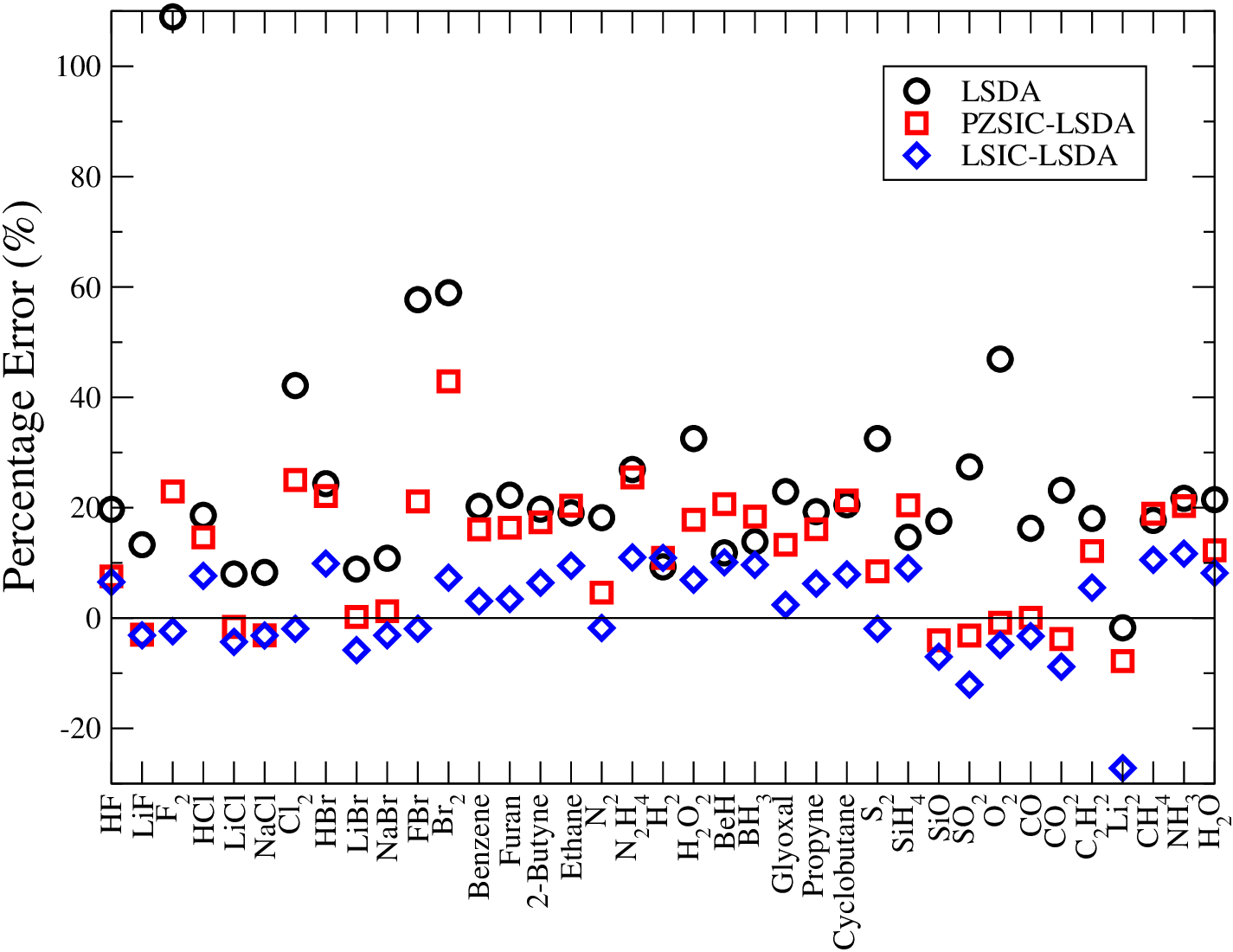}
\caption{\label{figatomization} Percentage errors for calculated atomization energies relative to experimental values.  Results for the LSDA, PZSIC-LSDA, and LSIC-LSDA methods are shown.}
\end{figure}

The percentage errors in calculated $AE$s relative to experiment are shown in Fig. \ref{figatomization} for LSDA, PZSIC-LSDA, and LSIC-LSDA.
LSDA significantly overestimates the $AE$s.  PZSIC-LSDA tends to improve them, but in most cases still overestimates their values.  LSIC-LSDA reduces the $AE$s further, bringing them into better agreement with experiment.
Mean absolute percentage errors (MAPE) for a variety of methods are compared in Table \ref{tabatomization}.  The MAPE for the full set of molecules is 24.21 \% for LSDA, 
13.42 \% for PZSIC-LSDA, and 6.94 \% for LSIC-LSDA. The performance of the LSIC-LSDA
falls between that of PBE (8.64 \%) and SCAN (5.22 \%).
We also list results for the AE6 test set in Table \ref{tabatomization}, showing both the MAE and MAPE. The AE6 molecules are intended to give a good representative of atomization energy performance.  Here, too, it can be seen that LSIC-LSDA has a performance that is better than PBE, though not as good as SCAN.

\begin{table*}
\caption{\label{tabatomization}Atomization energies: AE6 errors (MAE and MAPE) and errors for the full set (MAPE) are shown.}
\begin{ruledtabular}
\begin{tabular}{lccc}
Method      & AE6 MAE (kcal/mol)	& AE6 MAPE (\%) & 37 molecules MAPE (\%) \\ \hline
LSDA        &  74.26 &   15.93  & 24.21 \\
PBE	        & 13.43 & 3.31 & 8.64 \\
SCAN	    & 2.85  & 1.15 & 5.22 \\
PZSIC-LSDA	&  57.97 &  9.37 & 13.42 \\
PZSIC-PBE	&  18.83 &  6.82 & 9.67 \\
PZSIC-SCAN	&  16.31 &  5.64 & 10.24 \\
LSIC-LSDA	& 9.95  &  3.20  & 6.94 \\     		
\end{tabular}
\end{ruledtabular}
\end{table*}

%%%%%%%%%%%%%%%%%%%%%%%%%%%%%%%%%%%%%%%%%%%%%%%%%%%%%%%%%%%%%%%%%%%%
\subsection{SIE Test Sets} 
Recently, Sharkas \emph{et al.} \cite{doi:10.1021/acs.jpca.8b09940} used the FLO-SIC methodology to study the performance of the PZSIC for a test set consisting of dissociation energies \cite{C7CP04913G} (SIE4$\times$4) and reaction energies \cite{doi:10.1021/ct900489g} (SIE11) that are expected to be strongly affected by self-interaction errors in DFAs. They found that PZSIC significantly decreases the errors of LSDA and PBE relative to reference calculations using the coupled-cluster with singles, doubles and perturbative triple excitations (CCSD(T)) method. 
We studied the same test sets using LSIC-LSDA.
The SIE4$\times$4 set consists of four symmetric dimer cations at four different dimer separations R relative to the respective equilibrium separations R$_e$: R/R$_e =$ 1, 1.25, 1.5, and 1.75. The SIE11 set consists of six cationic reactions (of which four are the dimer cations from the SIE4$\times$4 data set at their equilibrium geometries) and five neutral reactions. 
We use the atomic geometries and FOD positions found in the supplementary information of Ref. [\citenum{doi:10.1021/acs.jpca.8b09940}] as starting points for our FLO-SIC-LSDA calculations. We re-optimized the FOD positions to ensure the FOD forces were below the $10^{-3}$ Ha/Bohr threshold.

Results for the individual cases in the test sets are shown in Table \ref{tab:dissociation} for LSDA, PZSIC-LSDA, and LSIC-LSDA. Results are given as signed errors (in kcal/mol) relative to accurate reference values (also shown).  For the SIE4$\times$4 case, PZSIC-LSDA and LSIC-LSDA clearly improve on the results of LSDA for all separations.  The PZSIC results are generally better than LSIC for $R/R_e > 1$, though the differences are small compared to the scale of the self-interaction corrections.  Conversely, the LSIC results are consistently better near $R_e$. The mean average error (MAE) is slightly smaller in LSIC than in PZSIC, 2.6 versus 3.0 kcal/mol, respectively. 
For the SIE11 test set, the signed errors are typically somewhat smaller in LSIC than in PZSIC.  In the case of the dissociation of (CH$_3$)$_2$CO$^+$ there is a dramatic reduction in the signed error.  This drops the MAE for the SIE11 cationic reactions from 14.83 for PZSIC to 2.31 kcal/mol for LSIC. The MAE for the SIE11 neutral reactions also shows an improvement from 9.01 to 6.31 kcal/mol.  

The LSIC-LSDA results are also as good or better than PZSIC-PBE results, which have MAE of 2.3, 8.7 and 7.9 kcal/mol for SIE4$\times$4, SIE11 cationic,  and SIE11 neutral, respectively. 

\begin{figure}
(a)\\
\includegraphics[width=\columnwidth]{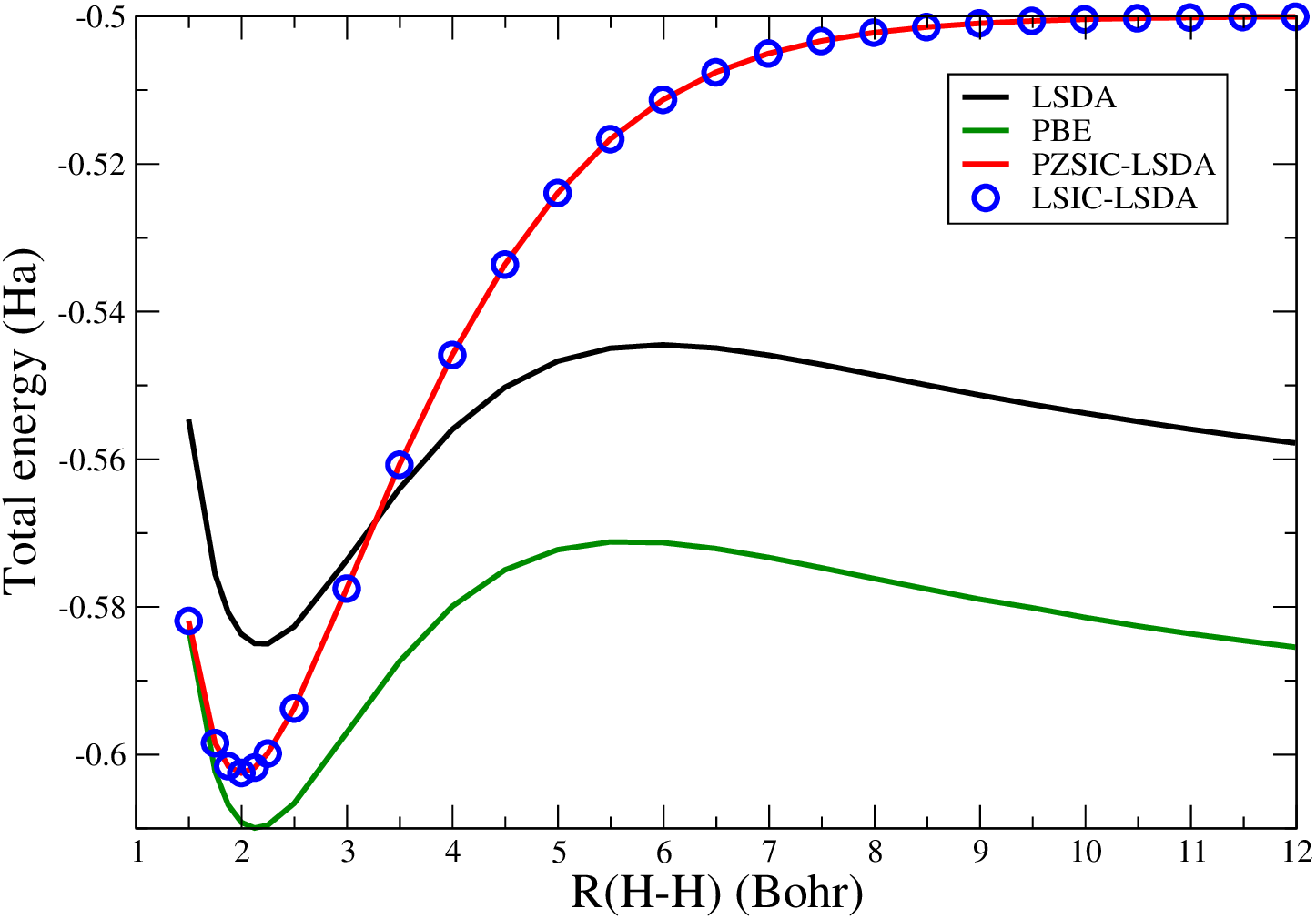}\\\vspace{7.5mm}
(b)\\
\includegraphics[width=\columnwidth]{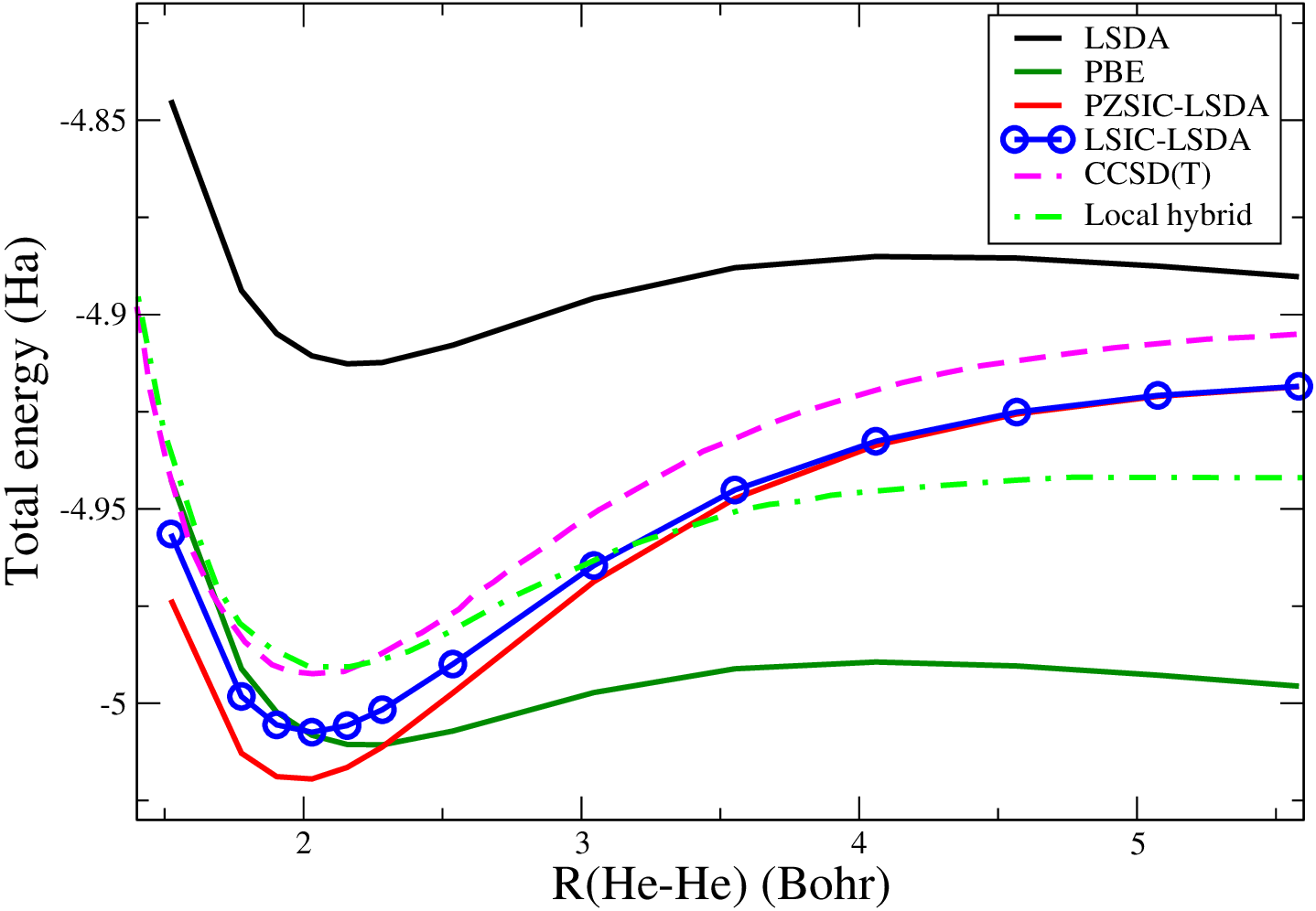}
\caption{\label{fig:dissociationcurve} Ground-state dissociation curves for (a) H$_2^+$ and (b) He$_2^+$. The CCSD(T)/cc-pVQZ results from Ref. [\onlinecite{doi:10.1063/1.2566637}] and local hybrid results from Ref. [\onlinecite{doi:10.1063/1.4865942}] are also shown for comparison.}
\end{figure}

Fig. \ref{fig:dissociationcurve} shows ground-state dissociation curves for H$_2^+$ and He$_2^+$. These give useful comparisons of the overall behavior of PZSIC-LSDA and LSIC-LSDA with that of LSDA and PBE both near and far from the equilibrium separations.  In both cases, the DFA calculations produce qualitatively incorrect energy curves as the interatomic separation increases, featuring a slight energy barrier at large separations 
on the way to a too-low energy for the dissociation products,
two H$^{+0.5}$ or He$^{+0.5}$ fragments.
Both PZSIC and LSIC calculations restore the correct qualitative shape to the dissociation curves.
In the case of H$_2^+$, PZSIC and LSIC give identical and exact results, because the iso-orbital indicator $z_{\sigma}$ is exactly one everywhere for this one-electron case.  For He$_2^+$, LSIC and PZSIC give the same results in the dissociation limit of He and He$^+$, since both of these are also one-electron systems (He has one-electron of each spin).  Near the equilibrium separation, however, LSIC reduces the size of the self-interaction correction, resulting in a binding energy that is close to that of PBE.

\begin{table*}
\caption{\label{tab:dissociation}
SIE4$\times$4 binding energy curves and SIE11 reaction energies
(kcal/mol). Reference energies and signed errors are shown.
}
\begin{ruledtabular}
\begin{tabular}{lcdcc}
\textrm{Reaction}&
\textrm{R/R$_e$}&
\multicolumn{1}{c}{\textrm{Ref.\footnote{Reference \cite{doi:10.1021/ct900489g}.}}}&
\textrm{PZSIC-LSDA\footnote{Reference \cite{FLOSICSCANpaper}.}}&
\textrm{LSIC-LSDA}\\
\colrule
H$_2^+ \rightarrow$ H $+$ H$^+$ & 1.0  & 64.4 &  0.0 & 0.0\\
                                & 1.25 & 58.9 &  0.0 & 0.0\\
                                & 1.5  & 48.7  &  0.0 & 0.0\\
                                & 1.75 & 38.3  &  0.1 & -0.1\\
\colrule
He$_2^+ \rightarrow$ He $+$ He$^+$ & 1.0 & 56.9 & 5.8 & -1.7\\   
                                   & 1.25 & 46.9  & 1.9 & -2.7\\
                                   & 1.5  & 31.3  & -0.4 & -3.0\\
                                   & 1.75 & 19.1  & -1.6 & -3.0\\
\colrule
(NH$_3)_2^+ \rightarrow$ NH$_3$ $+$ NH$_3^+$ & 1.0  & 35.9 & 11.7 & 1.6\\
                                             & 1.25 & 25.9  & 7.3 & 6.7\\
                                             & 1.5  & 13.4  & 4.0 & 8.1\\
                                             & 1.75 &  4.9 & 3.4  & 6.7\\
\colrule
(H$_2$O)$_2^+ \rightarrow$ H$_2$O $+$ H$_2$O$^+$ & 1.0  & 39.7 & 5.8 & 0.4\\
                                                 & 1.25 & 29.1  & -1.4 & 4.2\\
                                                 & 1.5  & 16.9  & -2.7 & 1.5\\
                                                 & 1.75 &  9.3  & -1.5 & 1.8\\

\colrule
\multicolumn{2}{l}{C$_4$H$_{10}^+ \rightarrow$ C$_2$H$_5 + $ C$_2$H$_5^+$ } & 35.28 & 11.44 & -6.00 \\
\multicolumn{2}{l}{(CH$_3)_2$CO$^+ \rightarrow$ CH$_3 +$ CH$_3$CO$^+$  } & 22.57 & 39.39 & 1.86\\
\multicolumn{2}{l}{ClFCl $\rightarrow$ ClClF  } & -1.01 & 4.63 & 4.37\\
\multicolumn{2}{l}{C$_2$H$_4$ \dots F$_2 \rightarrow$ C$_2$H$_4 +$ F$_2$ } & 1.08 & -0.23 & -2.82\\
\multicolumn{2}{l}{C$_6$H$_6$ \dots Li $\rightarrow$ Li $+$ C$_6$H$_6$ } & 9.50 & 10.19 & -13.50 \\
\multicolumn{2}{l}{NH$_3$ \dots ClF $\rightarrow$ NH$_3 +$ ClF } & 10.50 & 5.60 & -4.56\\
\multicolumn{2}{l}{NaOMg $\rightarrow$ MgO $+$ Na } & 69.56 & 28.56 & 11.45 \\
\multicolumn{2}{l}{FLiF $\rightarrow$ Li $+$ F$_2$ } & 94.36 & -4.82 & -1.18\\
\colrule
\multicolumn{2}{l}{MAE SIE4$\times$4    }&& 3.0 & 2.6 \\ 
\multicolumn{2}{l}{MAE SIE11 cationic  }&& 14.83 & 2.31\\
\multicolumn{2}{l}{MAE SIE11 neutral   }&& 9.01 & 6.31\\
\end{tabular}
\end{ruledtabular}
\end{table*}

%%%%%%%%%%%%%%%%%%%%%%%%%%%%%%%%%%%%%%%%%%%%%%%%%%%%%%%%%%%
\subsection{Barrier heights of chemical reactions}
To investigate the performance of LSIC for barrier heights in chemical reactions, we used the BH6 test set.  This is  a representative subset of the BH24 set \cite{doi:10.1021/ct600281g}. The reactions included in BH6 are:
OH + CH$_4 \rightarrow$ CH$_3$ + H$_2$O, H + OH $\rightarrow$ H$_2$ + O, and H + H$_2$S $\rightarrow$ H$_2$ + HS. Total energies at the left hand side, the right hand side, and the saddle point of these chemical reactions were evaluated, and the barrier heights of the forward (f) and reverse (r) reactions obtained by taking the relevant energy differences. 
We used the geometries provided in Ref. [\citenum{doi:10.1021/ct600281g}] and reference values from Ref. [\citenum{doi:10.1021/jp035287b}].
The results for various methods are summarized in Table \ref{tab:BH6}.

DFAs such as LSDA, PBE, and SCAN 
underestimate barrier heights \cite{doi:10.1063/1.2176608} by giving transition state energies that are too low compared to the reactant and 
product energies. An accurate description of the stretched bonds in the transition states
requires full nonlocality in the exchange-correlation potential that the semi-local functionals cannot provide.  Use of PZSIC reduces the overall errors, but in PZSIC-LSDA the barriers are still too small compared to reference values. 
This can be seen in the negative signed errors of all six barrier heights in Table \ref{tab:BH6}.
Applying LSIC-LSDA improves the barrier heights in almost every case.  The MAE of the 
barrier heights improves from 17.6 kcal/mol for LSDA, to 4.9 in PZSIC-LSDA, to only 1.3 kcal/mol in LSIC-LSDA.  Remarkably, as seen in the table, LSIC-LSDA has a smaller MAE than any of the methods listed, including PZSIC-PBE and PZSIC-SCAN.

\begin{table*}
\caption{\label{tab:BH6} BH6 forward (f) and reverse (r) barrier heights (kcal/mol). Signed errors are shown.}
\footnotetext{Reference \cite{doi:10.1021/jp035287b}.}
\begin{ruledtabular}
\begin{tabular}{cccccccccc}
 &&& \multicolumn{3}{c}{DFA} & \multicolumn{3}{c}{PZSIC} & LSIC \\ \cline{4-6} \cline{7-9} \cline{10-10}
 Reaction & Barrier & Ref.$^\text{a}$ & LSDA & PBE & SCAN & LSDA & PBE & SCAN & LSDA  \\\hline
  OH + CH$_4 \rightarrow$ CH$_3$ + H$_2$O & f & 6.7 & -23.6 & -12.2 & -8.3 & -2.2 & 5.7 & 4.6 & 2.6 \\
                                         & r & 19.6 & -17.4 & -10.7 & -7.8 & -12.5 & -10.3 & -7.1 & -0.2 \\
 H + OH $\rightarrow$ H$_2$ + O          & f & 10.7 & -11.8 & -2.2 & -7.5 & -1.1 & 2.3 & 0.0 & -0.6  \\
                                         & r & 13.1 & -25.3 & -9.9 & -11.0 & -4.8 & 2.9 & 1.8 & 1.2  \\       
 H + H$_2$S $\rightarrow$ H$_2$ + HS     & f & 3.6  & -10.3 & -4.8 & -6.3 & -1.7 & 1.7 & -1.9 & -1.3  \\ 
                                         & r & 17.3 & -17.2 & -8.1 & -6.2 & -7.0 & -2.1 & -2.2 & 2.2  \\\hline
 ME  & & & -17.6 & -8.0 & -7.9 & -4.9 &  0.0 & -0.8 & 0.7 \\
 MAE & & &  17.6 & 8.0  & 7.9 & 4.9 & 4.2 & 3.0 & 1.3 \\
\end{tabular}
\end{ruledtabular}
\end{table*}

%%%%%%%%%%%%%%%%%%%%%%%%%%%%%%%%%%%%%%%%%%%%%%%%%%%%%%%%%%%%%%%%%%
\section{\label{sec:discussion}Discussion}
 The LSIC method uses a point-wise scaling of SIC terms (Eqs. (\ref{eq:LSIC-DFA})--(\ref{eq:lsicxc})) to reduce the effect of self-interaction in many-electron regions, 
 while applying SIC at full strength in one-electron regions.
 We can think of this as interior orbital scaling, in comparison with the exterior orbital scaling of Eq. (\ref{eq:scalingscuseria}).  We showed in the previous section that using LSIC-LSDA results in significant performance gains for all the common electronic properties tested, as compared to both LSDA and PZSIC-LSDA. LSIC-LSDA improves on PZSIC for barrier heights and the SIE test sets where SIC is critical for getting physically reasonable results.  For near-equilibrium properties where PZSIC degrades the performance of semilocal DFAs, LSIC-LSDA gives results that are better than PBE and nearly as good as SCAN.  This is remarkable, given the relative simplicity of LSDA compared to the semilocal functionals. It is worth comparing and contrasting LSIC-LSDA results with results using an exterior orbital scaling method, such as that presented in Ref. [\citenum{doi:10.1063/1.2176608}].  These authors suggest $k =2$ as the best overall choice for use in Eq. (\ref{eq:factorscuseria}) and (\ref{eq:scalingscuseria}).  With this choice, the exterior orbital scaling method used with LSDA gives a MAE of 8.6 kcal/mol for the AE6 atomization energies.  This is slightly better than the LSIC-LSDA result of 9.95 kcal/mol shown in Table \ref{tabatomization}. For the BH6 barrier heights, the global scaling method gives a MAE of 4.7 kcal/mol, compared to 1.3 kcal/mol for LSIC-LSDA (Table \ref{tab:BH6}). While these results are similar, one should note that the global scaling approach causes the asymptotic form of the one electron potential to differ from the $-1/r$ form expected for the exact functional and maintained by PZSIC. This has an impact on properties that are sensitive to the nature of the potential in this region.  For the HOMO eigenvalues of the atoms H--Kr, for example, our investigation shows that the MAE for PZSIC-LSDA is 0.672 eV, and that for the orbital-wise scaling approach of Eq. (\ref{eq:factorscuseria}) and Eq. (\ref{eq:scalingscuseria}) is 1.034 eV ($k=1$) when compared to the experimental IPs. 
 Equality of the HOMO eigenvalue and the IP is a consequence of the linear variation of the total energy between adjacent integer numbers of electrons.  This many-electron self-interaction freedom \cite{doi:10.1063/1.2566637} is exact for the exact functional and approximately true for PZSIC.  It has been argued elsewhere \cite{doi:10.1063/1.2566637} 
that this property requires a full Hartree self-interaction correction term and thus should not be true for exterior orbital scaling corrections (or even for LSIC).  A similar problem involves dissociating heteroatomic molecules to the correct neutral atom limits\cite{doi:10.1063/1.2566637}. LSDA and the exterior orbital scaling method fail to do this in many cases, while PZSIC-LSDA succeeds. It is not yet clear how point-wise local scaling will affect such properties in general, since examining this requires fully self-consistent LSIC calculations. Preliminary {\it quasi} self-consistent LSIC calculations on the atomic systems using the weighted average of 
SIC potentials show that the self-consistency in fact slightly improves the LSIC results.
The MAE in the HOMO eigenvalues of {\it quasi} self-consistent LSIC results is 0.363 eV, compared to the 0.672 eV of perturbative LSIC and 1.034 of exterior orbital scaling (cf. Eq. (\ref{eq:scalingscuseria})). A full self-consistent implementation
of LSIC-LSDA has been formulated and is being implemented into the FLOSIC code.

\begin{figure}
\includegraphics[width=\columnwidth]{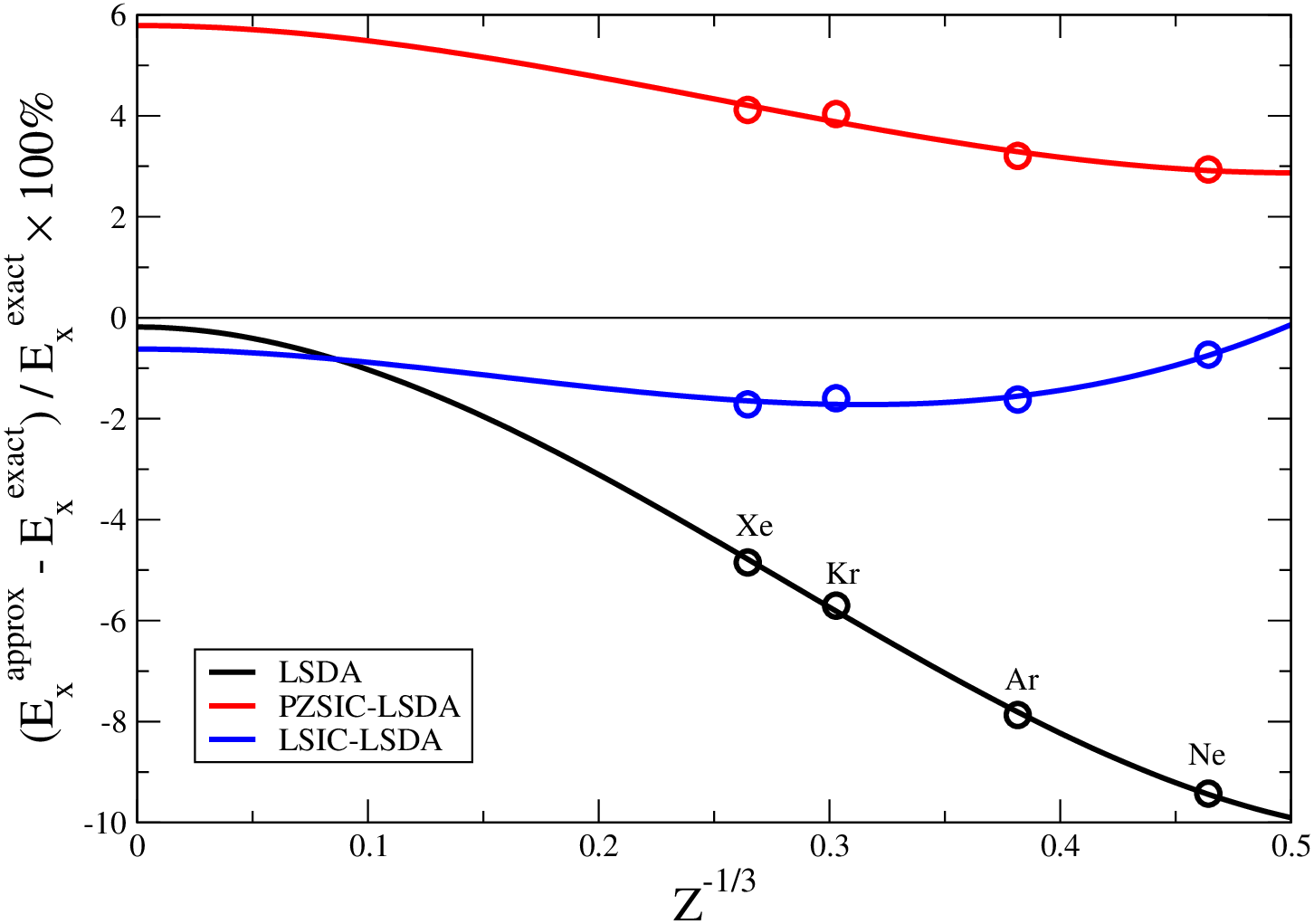}
\caption{\label{fig:exfitting} Percentage errors of the approximated exchange energies using the exact exchange energies as a reference. In LSIC-LSDA, Z$^{-1/3}$ $\rightarrow$ 0 limit is dramatically improved over PZSIC-LSDA.
}
\end{figure}

Recently, Santra and Perdew\cite{doi:10.1063/1.5090534} showed that uniform electron gas norms satisfied by  semi-local functionals are violated by the corresponding PZSIC-DFAs. To show how functionals behave in this limit, they fitted the calculated results for the exchange energy for neutral noble gas atoms using an exact large Z expansion of $E_X$ as a fitting function.
We used the same approach to test LSIC.  
We computed the LSIC-LSDA exchange energy of Ne, Ar, Kr, and Xe and then fitted these energies using the function,
\begin{equation}
    \frac{E_X^{approx}-E_X^{exact}}{E_X^{exact}} \times 100 \% = a + b x^2 + c x^3
\end{equation}
where $x=Z^{-1/3}$ and $a$, $b$, and $c$ are fit parameters.
The result is shown in Fig. \ref{fig:exfitting}.
The value of $a$ corresponds to the limit where $Z^{-1/3} \rightarrow 0$ which corresponds to the uniform density limit.  $a$ should vanish for the non-empircal LSDA, PBE, and SCAN functionals that are exact in this limit.
The reported values of $a$ are $-0.18$, $-0.06$, and $-0.28$ for LSDA, PBE, and SCAN and $5.79$, $-3.30$, and $-3.63$ for PZSIC-LSDA, PZSIC-PBE, and PZSIC-SCAN \cite{doi:10.1063/1.5090534}.  The small residual values of $a$ for LSDA, PBE, and SCAN are due to errors in the extrapolations.  For LSIC-LSDA, we obtain $a = -0.62 $.  We note that the scaling factor $z_\sigma(\vec{r}) = \tau_\sigma^W(\vec{r})/\tau_\sigma(\vec{r})$ we have chosen vanishes for a uniform density and that LSIC would therefore give no correction to LSDA in this limit. This may not be the case for a different choice of scaling factor. A constraint of the exact functional that is lost in PZSIC is thus recovered by LSIC (as by the exterior orbital scaling approach of Ref. [\citenum{doi:10.1063/1.2176608}]).   

\section{Conclusion}
We introduced the LSIC-LSDA method that incorporates a point-wise scaling of self-interaction corrections based on a simple
iso-orbital descriptor $z_{\sigma} = \tau_\sigma^W(\vec{r})/\tau_\sigma(\vec{r})$.  The 
essential idea is to retain the benefit of PZSIC in the regions where the self-interaction is expected to be strong, while reducing its effect in other regions.  
We showed the results of LSIC-LSDA for a number of properties, including  atomic total energies, IPs, and EAs for the atoms up to Kr, and  atomization energies for a subset of G2 molecules and  the AE6 molecules,  dissociation  and reaction energies of the SIE4$\times$4 and SIE11 test sets, and chemical reaction barriers for the BH6 data set.  
In nearly all cases, the performance of LSIC is dramatically improved compared to both pure LSDA and PZSIC-LSDA.
LSIC-LSDA even performs better than PBE for atomization energies and is competitive with SCAN in many cases,  while keeping the benefits of PZSIC for properties like barrier heights, where the semi-local functions do poorly. 
We also showed that LSIC-LSDA restores the correct uniform density limit of the 
exchange energy that is lost in PZSIC.  In all, LSIC-LSDA brings the full non-locality of the PZSIC 
method to bear for cases like stretched bonds where SIE effects are dominant, while maintaining the already good description 
of near equilibrium properties provided by semi-local functionals.

It is 
interesting to compare LSIC-LSDA with advancements made on the well-trodden path of creating and correcting more sophisticated semilocal functionals \cite{Pederson_5scientific}. A major
development along the latter was the introduction of a fraction of Hartree-Fock
exchange which resulted in mitigating many deficiencies of the pure density 
functional approaches. As mentioned earlier, 
the formal justification for such mixing was the adiabatic connection
between real interacting system and the non-interacting KS system. 
Because the exact exchange-correlation energy is an 
integral over coupling constant from 0 to 1, it could include some fraction of exact exchange, which is the correct 
integrand at the limit of zero coupling constant. 
It is interesting to see the parallels between 
this path and SIC.  
Because typical real systems are 
part-way between slowly-varying density and one-electron density
limits, the exact exchange-correlation energy could include some 
fraction of PZSIC, which is exact for any one-electron density. 
The 50\% scaling approach used by J\'onsson and coworkers \cite{doi:10.1063/1.4752229}
can be considered as a global (orbital-independent) hybrid of DFA
and PZSIC-DFA, in analogy to the traditional global hybrids.
On the other hand, the present LSIC approach
is analogous to  local hybrids.
Understanding obtained in the
development of hybrid  functionals may be beneficial in the further
development of the LSIC method.

\begin{acknowledgments}

The authors dedicate this paper to Dr. Mark Pederson
on the occasion of his 60th birthday. R.R.Z. and Y.Y. acknowledge Drs. Luis Basurto and Jorge Vargas for discussions and technical assistance.
This work was supported by the US Department of Energy, Office of 
Science, Office of Basic Energy Sciences, as part of the 
Computational Chemical Sciences Program under Award No. 
DE-SC0018331. The work of R.R.Z. was supported in part by
the US Department of Energy, Office of Science, 
Office of Basic Energy Sciences, under Award No. DE-SC0006818.
Support for computational time at the Texas Advanced 
Computing Center through NSF Grant No. TG-DMR090071, 
and at NERSC is gratefully acknowledged.
R.R.Z. and Y.Y. conceived the LSIC concept and prepared the first draft of the manuscript. The other authors contributed calculations, figures,
tables, references, discussions, and revision of the manuscript.
\end{acknowledgments}

\bibliography{bibtex_references} 

\end{document}